\documentclass[conference]{IEEEtran}

\usepackage{balance}
\usepackage{url}
\usepackage{graphicx}
\usepackage{subcaption}
\usepackage{float}

\usepackage{booktabs}
\usepackage{multirow}
\usepackage{pifont}

\usepackage{import}
\usepackage{tikz}
\usetikzlibrary{matrix, fit, shapes.geometric}

\usepackage{pgf}

\usepackage[]{siunitx}

\graphicspath{{figs/}}

\begin{document}

\date{}

\title{A DTLS Abstraction Layer for the Recursive Networking Architecture in RIOT}

\author{\IEEEauthorblockN{M. Aiman Ismail, Thomas C. Schmidt}
\IEEEauthorblockA{Internet Technologies Group, Dept. Informatik, HAW Hamburg, Germany\\
\{muhammadaimanbin.ismail, t.schmidt\}@haw-hamburg.de}}

\maketitle

\begin{abstract}
On the Internet of Things (IoT), devices continuously communicate with each other, with a gateway, or other Internet nodes.  Often devices are constrained and use insecure
channels for their communication, which exposes them to a selection of
attacks that may extract sensitive pieces of information or
manipulate dialogues for the purpose of sabotaging.

This paper presents a new layer in the RIOT networking
architecture to seamlessly integrate  secure communication between applications using DTLS.
The layer acts as a modular abstraction layer of the different DTLS
implementations, enabling swapping of the underlying implementation with just
a few lines of code. This paper also introduces credman, a new module to manage
credentials used for (D)TLS connections.
\end{abstract}

\section{Introduction}

Security is an important part when communicating through the Internet.
Despite the fact that without proper security practices, bad actors could break into our network
infrastructures and cause severe damage to parties involved,
there are still numerous devices, IoT appliances in particular, that expose themselves on
the Internet without having any proper security measures in place.

Datagram Transport Layer Security (DTLS) \cite{RFC-6347} is a protocol for traffic encryption on top of UDP \cite{RFC-768}.
It is based on the concepts of TLS \cite{RFC-8446} and provides equivalent
security guarantees. DTLS guarantees reliable transport during the handshake
process but maintains UDP transport properties during application data transfer.
The protocol is deliberately designed to be as similar to TLS as possible, both to
minimize new security inventions and to maximize the amount of code and
infrastructure reuse.

RIOT \cite{bghkl-rosos-18} is an open source real-time OS, based on a modular architecture built around
a lightweight micro-kernel, and developed by a worldwide community of developers.
The modular approach enables easy prototyping and development to test new ideas
and deploy applications. Its default network stack GNRC follows a cleanly layered,
recursive design that easily allows for stacking and exchanging protocol layers or implementations.

In this paper, we describe how we built the DTLS abstraction layer on top of
existing components in the RIOT networking architecture. This layer provides an API
that can be implemented using third-party DTLS libraries. It is designed to be
independent of the underlying DTLS implementation, therefore allows the DTLS
stack to be exchanged without altering the applications that uses it. We also introduce
a new RIOT module \textit{credman} to manage the credentials used for the handshake.

The remainder of this paper is structured as follows. In Section II, we introduce the
existing networking stack of RIOT. In Sections III, we describe  the new secure
network stack, and Section IV presents experiments that assess its performance. In Section V, we draw
conclusions with an outlook on future work.

\section{RIOT networking subsystem}

\begin{figure}
  \centering
  \includegraphics[width=\columnwidth]{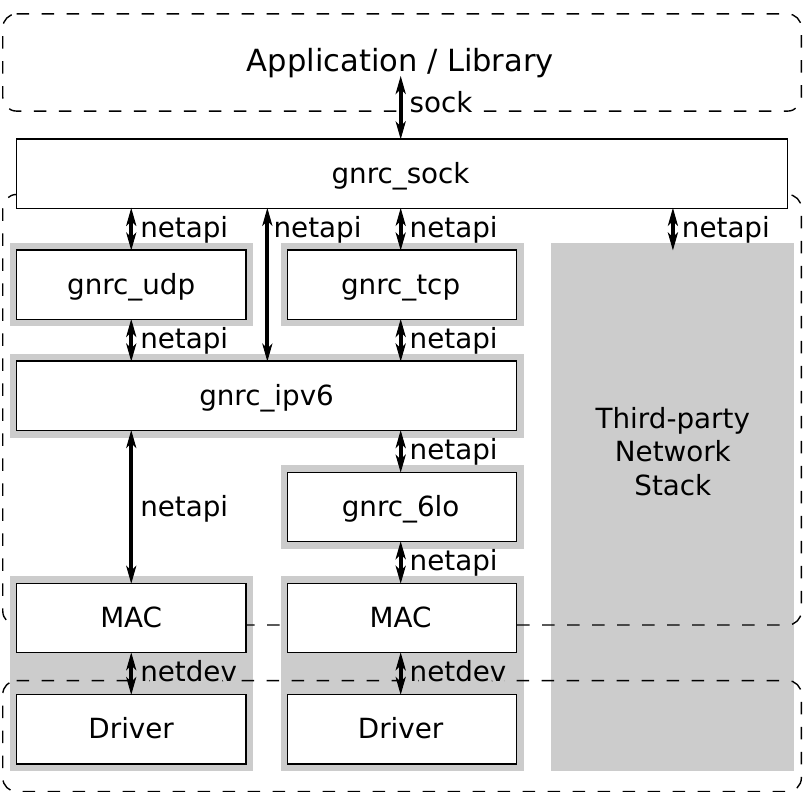}
  \caption{RIOT networking stack  }
  \label{fig:riot-net}
\end{figure}

The RIOT networking subsystem is designed to follow a modular architecture with clean
interfaces for abstracting all layers \cite{lkhpg-cwemr-18}. This facilitates the creation
and integration of new protocols, different implementations, or additional layers such as
a new encryption layer to the existing stack. It consists of the two external
APIs \textit{netdev, sock}, and a single internal API for communication between layers,
\textit{netapi}. It is noteworthy that the RIOT networking subsystem simultaneously
supports multiple interfaces with different protocol stacks, which makes it capable of
running gateway services.  An architectural overview is visualized in Figure \ref{fig:riot-net}.

\textit{The Device Driver API: netdev}. Individual network devices in RIOT are
abstracted via \textit{netdev}, which allows networking stacks access to the
devices via a common, portable interface. \textit{netdev} remains neutral
in that it does not enforce implementation details regarding memory allocation,
data flattening, and threading. These decisions are delegated to the users
of the interface.

\textit{The Internal Protocol Interface: netapi}. Internal protocol
layers in the RIOT networking subsystem can be recursively composed via the \textit{netapi}.
The interface is kept simple so that even an exotic networking protocol could
be implemented against it. Message passed between layers are typed
as following: two asynchronous message types
(\texttt{MSG\_TYPE\_SND, MSG\_TYPE\_RCV}) and two synchronous message types
(\texttt{MSG\_TYPE\_GET, MSG\_TYPE\_SET}) that expects a reply in form of
\texttt{MSG\_TYPE\_ACK} typed message. No further semantic are built into
the messages of \textit{netapi}, but certain preconditions on packets or
option values handed to \textit{netapi} can be set as requirements to
implement more complex behavior that goes beyond these plain specification.

\textit{The User Programming API: sock}. This module provides a network API for
applications and libraries in RIOT. It provides a set of functions to establish
connections or send and receive datagrams using different types of protocols.
In comparison to POSIX sockets, \textit{sock} does not require complex and
memory expensive implementation and therefore more suited for use in
constrained hardware. Only common type and definitions from either
\textit{libc} or POSIX. This ensures that \textit{sock} is easy to port to
other target OS.

\textit{GNRC} is the native IPv6 networking stack for RIOT. It takes full advantage of the
multi-threading model supported by RIOT to foster a clean protocol separation
via well-defined interfaces and IPC. Each network protocol is encapsulated
in its own thread and uses RIOT thread-targeted IPC with a message queue in each
thread to communicate between layers. Other stacks that introduce different networking
protocols such as ICN also integrate via the same interfaces. Various experimental
evaluations and benchmarks \cite{lkhpg-cwemr-18,gklp-ncmcm-18} have proven the feasibility
and efficiency of this flexible approach to networking in RIOT.

\section{Introducing the secure network stack}

\begin{figure}[b]
  \centering
  \begin{tikzpicture}[>=latex]
    \tikzset{
        cell/.style={
            rectangle, rounded corners=5pt, thick, draw,
        }
    }

    \matrix[
        matrix of nodes, nodes in empty cells,
        minimum width=3cm,
        column sep=6pt,
        row sep=6pt,
        minimum height=4ex,
    ] (arch) {
        & \\
        & \\
        & \\
        & \\
    };

    \node[cell, fit=(arch-4-1)(arch-4-1), inner sep=0pt, label=center: sock\_udp] {};
    \node[cell, fit=(arch-3-1)(arch-3-1), inner sep=0pt, label=center: DTLS Libraries] {};
    \node[cell, fit=(arch-4-2)(arch-3-2), inner sep=0pt, label=center: credman] {};
    \node[cell, fit=(arch-2-1)(arch-2-2), inner sep=0pt, label=center: sock\_dtls] {};
    \node[cell, fit=(arch-1-1)(arch-1-1), inner sep=0pt, label=center: Application] {};
    \node[cell, fit=(arch-1-2)(arch-1-2), inner sep=0pt, label=center: Application] {};
\end{tikzpicture}
  \caption{Architecture of sock\_dtls}
  \label{fig:arch}
\end{figure}
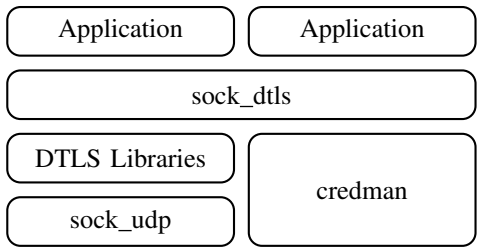

The modular nature of the existing GNRC stack allows for an easy extension by
adding DTLS at the top  while maintaining its modularity. We introduced
two new modules \textit{credman} and \textit{sock\_dtls} (see Figure \ref{fig:arch}).

\textit{credman} is a module to manage credentials used in (D)TLS encryption
protocols. Credentials registered with the system are identified by using the
tuple of int-based tag \textit{credman\_type\_t} and the credential type
\textit{credman\_tag\_t}. This in combination with the \textit{sock\_dtls} API
allows users to register multiple credentials of the same type, which can be the case
if the nodes are communicating with multiple other nodes simultaneously and each
node uses different credentials for authentication.

\textit{credman} does not copy the credentials into the system memory. It only
has information about the credentials and points to the location of the credential
itself, which can be stored in protected regions of the memory. Users will have to
ensure that a credential is available at the location given to credman during the
lifetime of their application.

We defined a new \textit{sock} type --- \textit{sock\_dtls}. It is designed
to mimic the behavior of \textit{sock\_udp} as closely as possible so that
integrating it into existing applications and libraries can be done without
introducing too many new changes. By adding a line in the Makefile, users can choose
which underlying DTLS implementation to use. Swapping to a new DTLS implementation
is simply done by specifying the corresponding implementation in the Makefile.
Through this mechanism, testing and evaluation of DTLS implementations can be
performed without altering the application.

Figure \ref{fig:arch} summarizes the integration of the DTLS abstraction layer with
existing network stack in RIOT. Currently, RIOT only has support for
tinyDTLS\footnote{\url{https://projects.eclipse.org/projects/iot.tinydtls}}
but there is ongoing work\footnote{\url{https://github.com/RIOT-OS/RIOT/pull/10308}} to add
support for wolfSSL\footnote{\url{https://www.wolfssl.com}}.

The use cases of \textit{sock\_dtls} are twofold, the DTLS server and the DTLS client.
The server is created by \texttt{sock\_dtls\_create()}. Then it needs to tell the
credentials to sock by \texttt{sock\_dtls\_register\_credential\_tags()}.
After the call \texttt{sock\_dtls\_init\_server()} the server is ready to receive new DTLS
session establishment requests from clients.

DTLS clients also need to create the sock using \texttt{sock\_dtls\_create()} and
then register the credentials with \texttt{sock\_dtls\_register\_credential\_tags()}.
After that, a session to a DTLS server can be established using
\texttt{sock\_dtls\_establish\_session()}. If successful, the session can
be used to send and receive datagram packet like in a normal UDP channel.

For the DTLS server, the operations can be summarised as follows:

\begin{enumerate}
  \item Create \textit{sock\_dtls}
  \item Register credentials available for use
  \item Initialize the server
  \item Start listening for incoming datagram packets
\end{enumerate}

As for the DTLS client, the first two steps are the same as for the server followed by

\begin{enumerate}
  \setcounter{enumi}{2}
  \item Establish session with a DTLS server
  \item Start sending and receiving datagram packets
\end{enumerate}

\section{Experiments and Evaluation}

\begin{figure}
  \includegraphics[width=0.94\columnwidth]{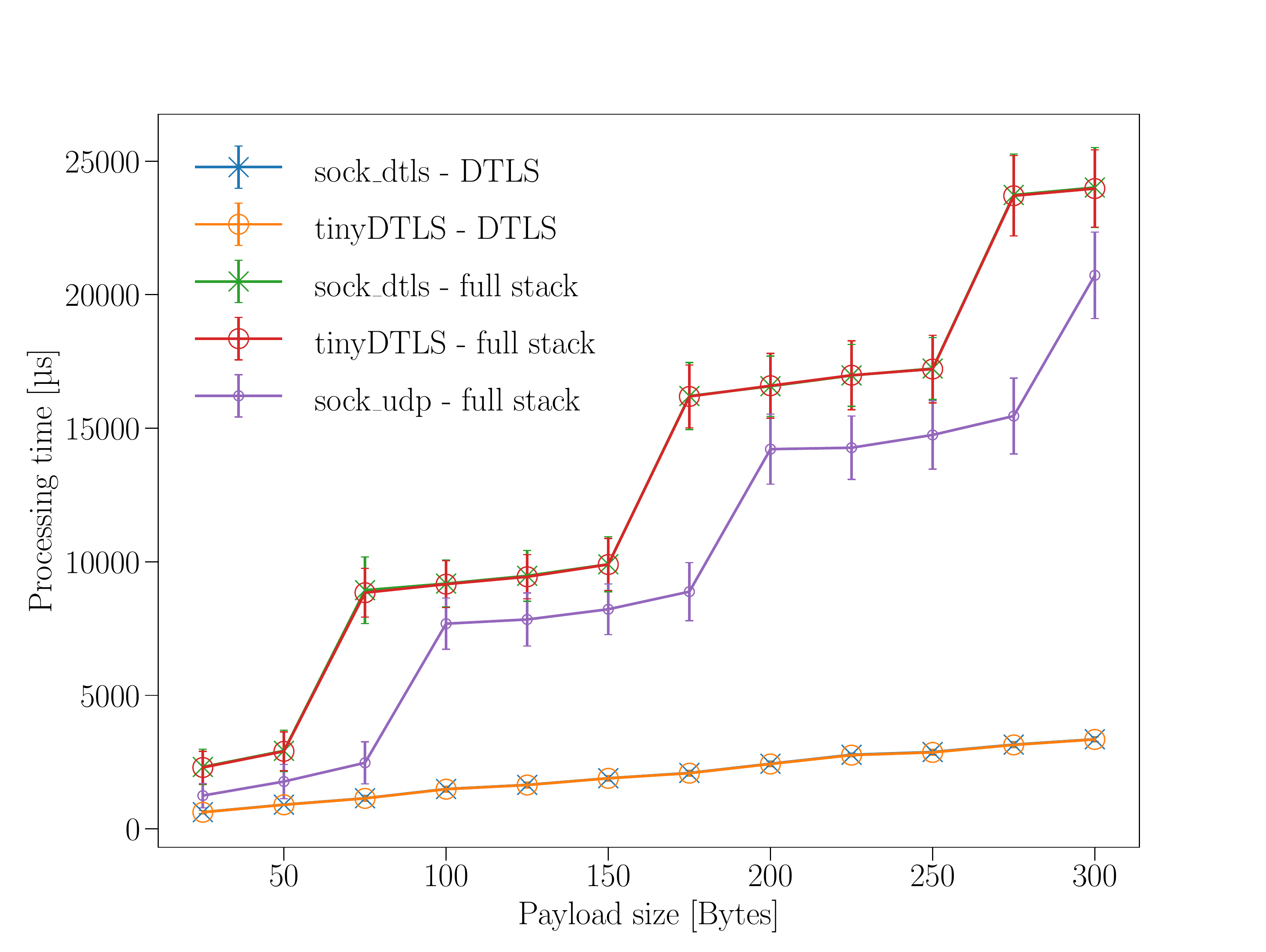}
  \caption{CPU Overhead}
  \label{fig:cpu-overhead}
\end{figure}

\begin{figure}
  \includegraphics[width=0.94\columnwidth]{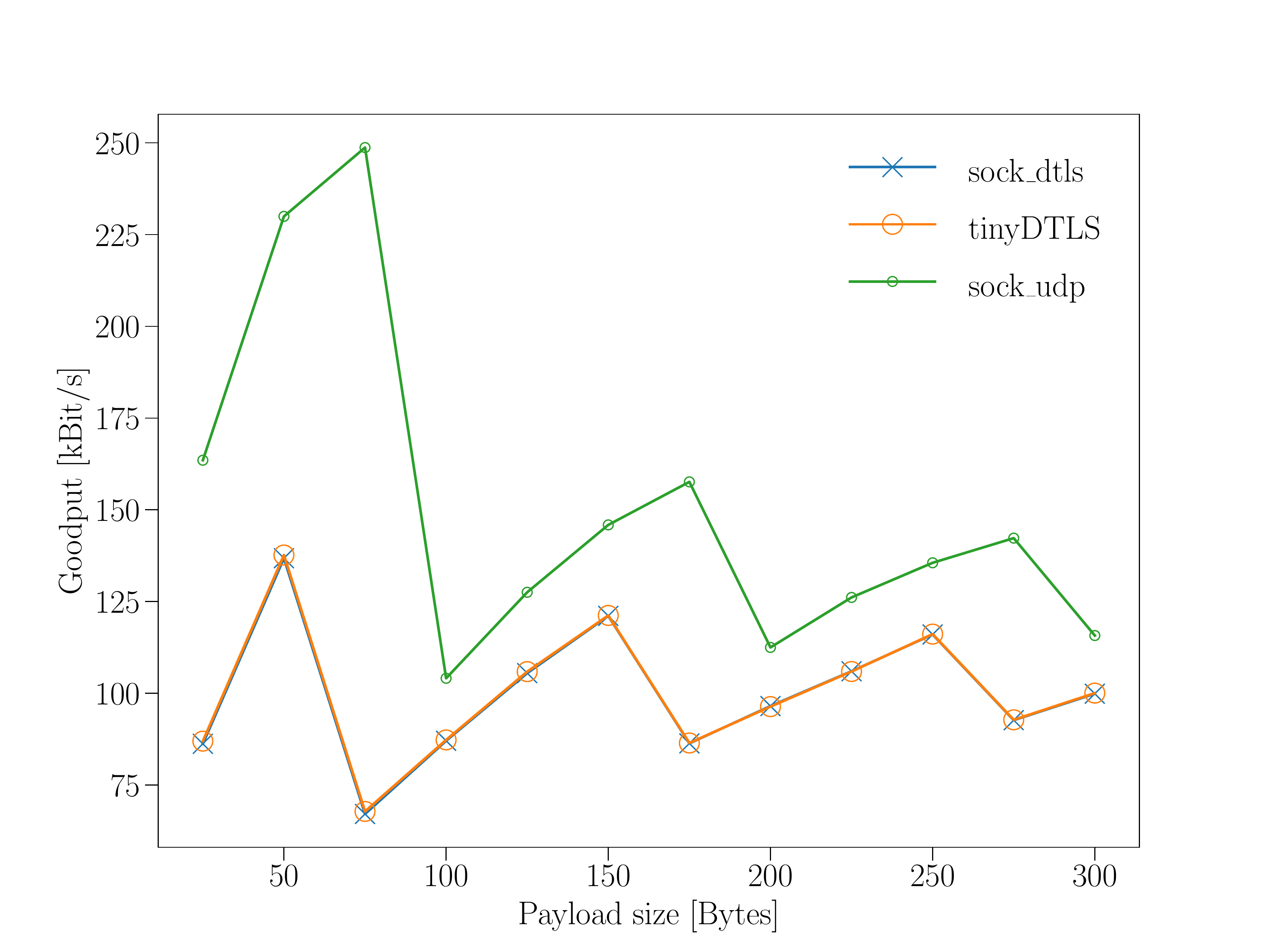}
  \caption{Average Goodput}
  \label{fig:goodput}
\end{figure}

\begin{figure}[t]
  \includegraphics[width=0.94\columnwidth]{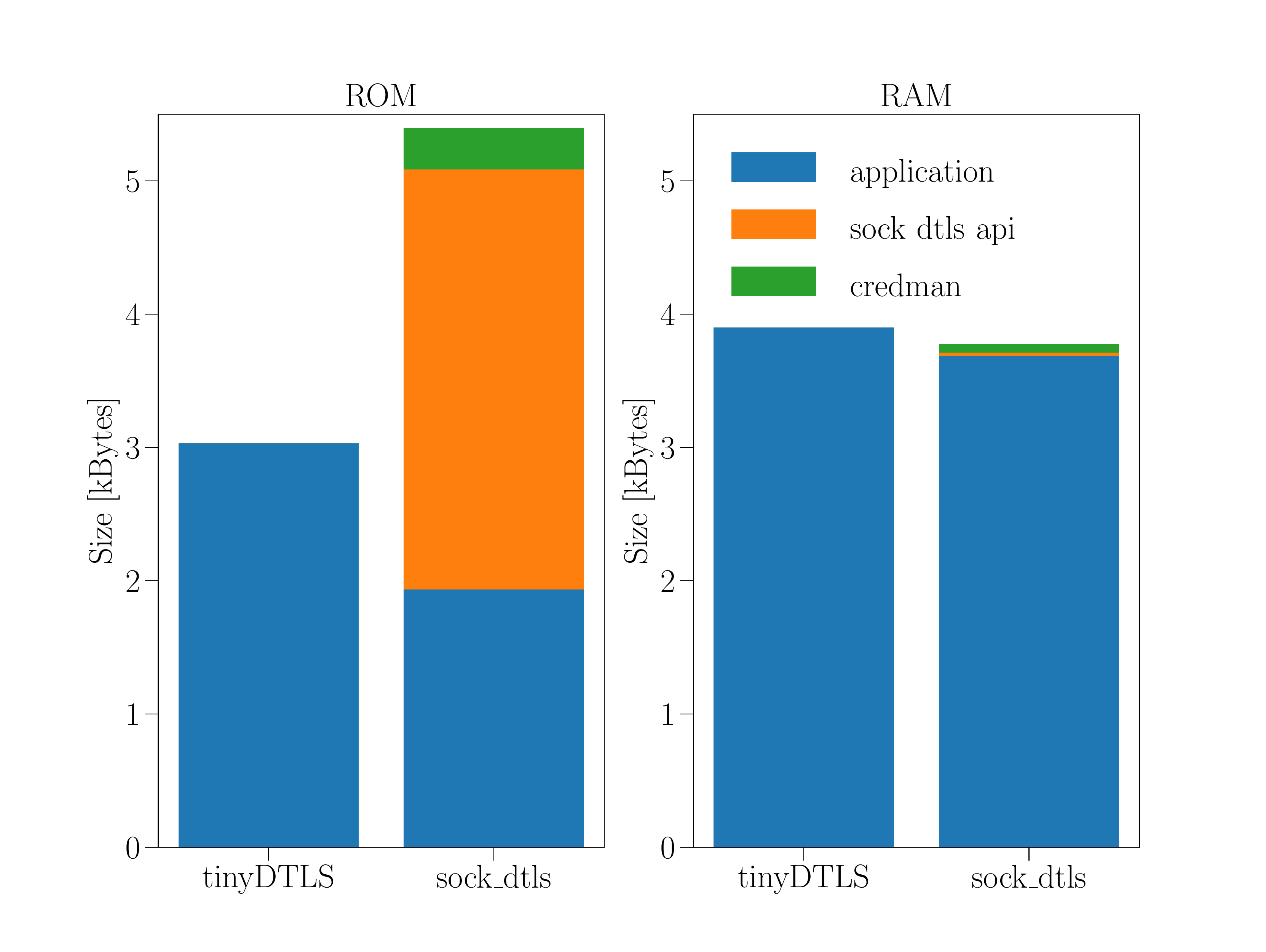}
  \caption{Memory Usage}
  \label{fig:memory}
\end{figure}

We are now ready to validate our concept and assess the performance of our implementation.
We compared \textit{sock\_dtls} with tinyDTLS and \textit{sock\_udp} and examine
the metrics CPU overhead and goodput during the transmission of payloads as well as
memory consumption. All measurements were
performed on  \textit{samr21-xpro} boards positioned side-by-side over the
802.15.4 wireless radio network \cite{IEEE-802.15.4-11} with 6LoWPAN encapsulation and
header compression \cite{RFC-4944}.

We wrote three versions of a client and a server program that send  packets
of  increasing payload sizes from the client to the server while recording the
time taken to transmit the packets. The first version uses tinyDTLS directly
while the second version uses our new DTLS abstraction layer \textit{sock\_dtls}
with tinyDTLS. The third version employs no encryption layer but plainly uses RIOT UDP
sock API \textit{sock\_udp} to transmit the packets and acts as a controlling baseline.

The test was setup as follows unless stated otherwise. The server is instantiated to listen for new connections
and receives packet from clients. For each received packet, the payload size is
logged into a file. On the client side, two metrics are measured: the time
taken to process a packet for the full network stack, that is (1) DTLS, UDP, IP, 6LoWPAN, MAC,
and auxiliary components, and (2) the time taken to process only the DTLS part of the
transmission, which starts from accepting the packet from user and encrypting it using
specified keying materials to just before passing it to UDP layer for further processing.
The test is run with payload size  ranging between 25 Bytes and 300 Bytes in 25 Bytes
 intervals. Each configuration is repeated 5000 times with averages and standard deviations
 recorded in the following diagrams.

\textbf{CPU Overhead}. Figure \ref{fig:cpu-overhead} depicts the CPU overhead
during packet transmission. The test program using \textit{sock\_dtls} and
tinyDTLS need approximately the same average processing time per packet with
\textit{sock\_dtls} being slightly higher. The extra overhead when adding an
abstraction layer is expected as a tradeoff for faster prototyping time and
ease of use, which in this case is virtually negligible. The step-like line
shaped for the full stack processing can be attributed to the fragmentation of packets by the underlying 6LoWPAN layer
 when certain size limits are reached. Comparison of the times taken to process
only the DTLS layer shows an almosts linear line of the processing time with
increasing payload size and again, there is only little difference between the values.

This clearly indicates that our  \textit{sock\_dtls} abstraction layer comes at negligible processing overhead.

\textbf{Goodput}. The average goodput is shown in  in Figure \ref{fig:goodput}. It follows the same trend with the
\textit{sock\_dtls} version admitting approximately the same performance values as the
tinyDTLS version. These results not only indicate a  picture consistent with processing, but also confirm the robustness of our interface layer.

\textbf{Memory}. Figure \ref{fig:memory} compares the memory consumptions of the different DTLS code versions.
Here we measured the memory usage of a simple echo client and server application implemented using \textit{sock\_dtls}
and tinyDTLS instead of our test application to mirror a more complete DTLS application compared to the test application.
The hardware setup is the same.

The RAM usage of both programs is similar with \textit{sock\_dtls} saving around 120 Bytes  compared to tinyDTLS.
This saving is mainly contributed by the compiler, which can optimize away some of the variables used for the sending and receiving functions in user application but not in tinyDTLS.
As a result, even though we actually need about 80 Bytes more in \textit{sock\_dtls} for \textit{credman} and the API,
we still end up using less RAM. Nevertheless, because the saving is only around 100 Bytes and is mainly caused by
compiler optimization, we could actually say that the RAM usage is approximately the same in both versions and the exact value
is determined by the quality of implementation in user application.

In contrast, the ROM usage in \textit{sock\_dtls} is about two kilobytes larger than in tinyDTLS.
The larger ROM size is due to the code size of \textit{sock\_dtls}.
This value is implementation specific as each implementation needs to be implemented
against the DTLS sock interface first before used as the underlying
implementation of \textit{sock\_dtls}. When using tinyDTLS specifically, we could
actually delegate the bulk of credential management  to \textit{credman}. For tinyDTLS this must be implemented as
callbacks by the users. This simplifies the user application and
actually achieves about the same performance using less code.

\section{Conclusion and Future Works}

In this paper, we  introduced and analyzed the new DTLS abstraction layer
designed to be modular and easy for integrating into existing applications.
We demonstrated that the tradeoff between performance and ease of use is well acceptable for
normal use cases. Leveraging a clean and implementation-independent interface,
we increased the portability of applications and also the maintainability of upper
layer protocol implementations such as CoAP \cite{RFC-7252} over time.

In the future, we will work on implementing a DTLS profile for authentication and
authorization for the constrained environment such as \cite{draft-gerdes-ace-dtls-authorize}
to provide a framework for a secure network infrastructure. The integration of \textit{sock\_dtls}
in upper layer protocols such as the RIOT \textit{gcoap}\footnotemark is also on our schedule.

\balance

\bibliographystyle{ieeetran}
\bibliography{own,rfcs,ids,layer2}

\begin{thebibliography}{10}
\providecommand{\url}[1]{#1}
\csname url@samestyle\endcsname
\providecommand{\newblock}{\relax}
\providecommand{\bibinfo}[2]{#2}
\providecommand{\BIBentrySTDinterwordspacing}{\spaceskip=0pt\relax}
\providecommand{\BIBentryALTinterwordstretchfactor}{4}
\providecommand{\BIBentryALTinterwordspacing}{\spaceskip=\fontdimen2\font plus
\BIBentryALTinterwordstretchfactor\fontdimen3\font minus
  \fontdimen4\font\relax}
\providecommand{\BIBforeignlanguage}[2]{{%
\expandafter\ifx\csname l@#1\endcsname\relax
\typeout{** WARNING: IEEEtran.bst: No hyphenation pattern has been}%
\typeout{** loaded for the language `#1'. Using the pattern for}%
\typeout{** the default language instead.}%
\else
\language=\csname l@#1\endcsname
\fi
#2}}
\providecommand{\BIBdecl}{\relax}
\BIBdecl

\bibitem{RFC-6347}
E.~Rescorla and N.~Modadugu, ``{Datagram Transport Layer Security Version
  1.2},'' IETF, RFC 6347, January 2012.

\bibitem{RFC-768}
J.~Postel, ``{User Datagram Protocol},'' IETF, RFC 768, August 1980.

\bibitem{RFC-8446}
E.~Rescorla, ``{The Transport Layer Security (TLS) Protocol Version 1.3},''
  IETF, RFC 8446, August 2018.

\bibitem{bghkl-rosos-18}
\BIBentryALTinterwordspacing
E.~Baccelli, C.~G{\"u}ndogan, O.~Hahm, P.~Kietzmann, M.~Lenders, H.~Petersen,
  K.~Schleiser, T.~C. Schmidt, and M.~W{\"a}hlisch, ``{RIOT: an Open Source
  Operating System for Low-end Embedded Devices in the IoT},'' \emph{IEEE
  Internet of Things Journal}, vol.~5, no.~6, pp. 4428--4440, December 2018.
  [Online]. Available: \url{http://dx.doi.org/10.1109/JIOT.2018.2815038}
\BIBentrySTDinterwordspacing

\bibitem{lkhpg-cwemr-18}
\BIBentryALTinterwordspacing
M.~Lenders, P.~Kietzmann, O.~Hahm, H.~Petersen, C.~G{\"u}ndogan, E.~Baccelli,
  K.~Schleiser, T.~C. Schmidt, and M.~W{\"a}hlisch, ``{Connecting the World of
  Embedded Mobiles: The RIOT Approach to Ubiquitous Networking for the Internet
  of Things},'' Open Archive: arXiv.org, Technical Report arXiv:1801.02833,
  January 2018. [Online]. Available: \url{https://arxiv.org/abs/1801.02833}
\BIBentrySTDinterwordspacing

\bibitem{gklp-ncmcm-18}
\BIBentryALTinterwordspacing
C.~G{\"u}ndogan, P.~Kietzmann, M.~Lenders, H.~Petersen, T.~C. Schmidt, and
  M.~W{\"a}hlisch, ``{NDN, CoAP, and MQTT: A Comparative Measurement Study in
  the IoT},'' in \emph{Proc. of 5th ACM Conference on Information-Centric
  Networking (ICN)}.\hskip 1em plus 0.5em minus 0.4em\relax New York, NY, USA:
  ACM, September 2018, pp. 159--171. [Online]. Available:
  \url{https://conferences.sigcomm.org/acm-icn/2018/proceedings/icn18-final46.pdf}
\BIBentrySTDinterwordspacing

\bibitem{IEEE-802.15.4-11}
{IEEE 802.15 Working Group}, ``{IEEE Standard for Local and metropolitan area
  networks---Part 15.4: Low-Rate Wireless Personal Area Networks (LR-WPANs)},''
  IEEE, New York, NY, USA, Tech. Rep. IEEE Std 802.15.4\texttrademark--2011,
  Sep 2011.

\bibitem{RFC-4944}
G.~Montenegro, N.~Kushalnagar, J.~Hui, and D.~Culler, ``{Transmission of IPv6
  Packets over IEEE 802.15.4 Networks},'' IETF, RFC 4944, September 2007.

\bibitem{RFC-7252}
Z.~Shelby, K.~Hartke, and C.~Bormann, ``{The Constrained Application Protocol
  (CoAP)},'' IETF, RFC 7252, June 2014.

\bibitem{draft-gerdes-ace-dtls-authorize}
S.~Gerdes, O.~Bergmann, C.~Bormann, G.~Selander, and L.~Seitz, ``{Datagram
  Transport Layer Security (DTLS) Profile for Authentication and Authorization
  for Constrained Environments (ACE)},'' IETF, Internet-Draft -- work in
  progress~01, March 2017.

\end{thebibliography}

\footnotetext{\url{https://riot-os.org/api/group__net__gcoap.html}}

\end{document}